# Bayesian Learning of Neural Networks for Signal/Background Discrimination in Particle Physics

Michael Pogwizd[*], Laura Jane Elgass[**], Pushpalatha C. Bhat[†]

*University of Illinois*
*\*\* College of DuPage*
*† Fermi National Accelerator Laboratory*

**Abstract.** Neural networks are used extensively in classification problems in particle physics research. Since the training of neural networks can be viewed as a problem of inference, Bayesian learning of neural networks can provide more optimal and robust results than conventional learning methods. We have investigated the use of Bayesian neural networks for signal/background discrimination in the search for second generation leptoquarks at the Tevatron, as an example. We present a comparison of the results obtained from the conventional training of feedforward neural networks and networks trained with Bayesian methods.

## INTRODUCTION

Multivariate methods can be used to improve several aspects of a high energy physics (HEP) data analysis [1]. Finding signals of new physics is one of the important tasks in high energy physics, and signal/background discrimination can benefit from the use of multivariate methods when correlations between variables exist. Feedforward neural networks (which we refer to as "conventional" neural networks) have become the most popular multivariate method employed in HEP data analysis because of their power and ease of use. They are now used quite extensively in particle physics.

## BAYESIAN LEARNING

The elegance and advantage of neural networks derive from the fact that when trained for binary classification (signal vs. background), the output of a feedforward neural network approximates the probability for an event to be signal (= p(s | data)). However, there are limitations to conventional neural networks. First, training of the neural network yields one set of weights or network parameters, which requires looking for the "best" network. This can result in over-fitting which needs to be avoided. Additionally, heuristic decisions are needed on the network architecture, such as inputs and the number of hidden nodes. Finally there is no direct way to compute uncertainties.

Bayesian neural networks [2] are modeled after Bayes' Theorem, and treat the problem of classification as a problem of inference. Bayesian learning of neural networks can be more optimal and robust than conventional neural networks. Bayesian training provides a posterior density for the network weights $w$ -- p($w$|training data). Generated in the network parameter space is a sequence of weights (network parameters) -- *i.e.* a sequence of networks. The optimal network is approximated by averaging over the last K networks. This decision by averaging over many networks has lower error than that of any individual network. The advantages, then, of Bayesian Learning are that the neural networks are less prone to over-fitting, there is less need to optimize the size of the network (because a low probability density will be assigned to points that correspond to unnecessarily large networks), and in principle, this provides the best estimate of p(s | data).

### Search for Second Generation Leptoquarks

The comparison of Bayesian and feedforward neural networks was performed using the search for second generation leptoquarks as an example. Leptoquarks are hypothesized particles that have both lepton and quark characteristics and would serve to explain the remarkable symmetry between leptons and quarks. Second generation leptoquarks (LQ) are produced in pairs at the Tevatron ($LQ, \overline{LQ}$). Each leptoquark decays into a muon and a quark (jet): $p\overline{p} \rightarrow LQ\overline{LQ} \rightarrow \mu j \mu j$. The current mass limit for LQ is 247 GeV- 251 GeV (D0). Main backgrounds come from the Drell-Yan, top pair production $t\overline{t}$, and WW + jets production. We used JetNet [3] package for conventional neural networks and Radford Neal's software to construct Bayesian Neural Networks [4].

Four variables -- dimuon mass $M(\mu\mu)$, ratio of computed mass difference between the two leptoquarks in the event normalized by the average mass $\Delta M(LQ)/\langle M(LQ) \rangle$, four body invariant mass $M(\mu j \mu j)$, and sum transverse energy $E_T$ from LQ (mass = 240 GeV; LQ240) and $t\bar{t}$ were used in the training and analysis of the neural networks (See Fig. 1).

**Figure 1.** Distributions of variables used for LQ of mass 240GeV and $t\bar{t}$.

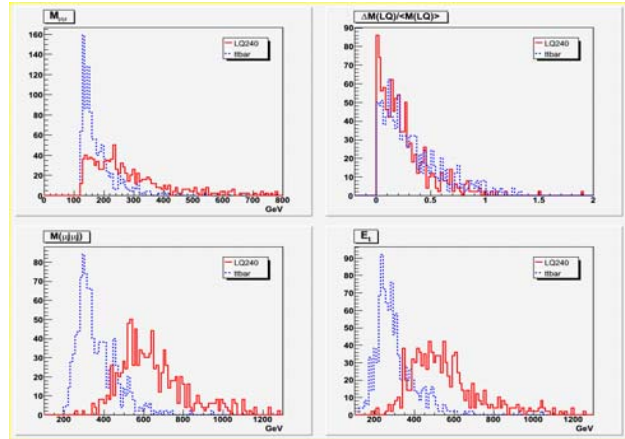

Both neural networks consisted of 4 input nodes, 5 hidden nodes, and one output node (set to 0 for background and 1 for signal, for training). Each network was trained with 1000 events for both signal (LQ240) and background ($t\bar{t}$) with 1000 iterations. The neural networks were then tested with the remaining signal and background events. The neural network outputs are shown in Fig. 2.

**Figure 2**. Output distributions for Conventional and Bayesian Neural Networks.

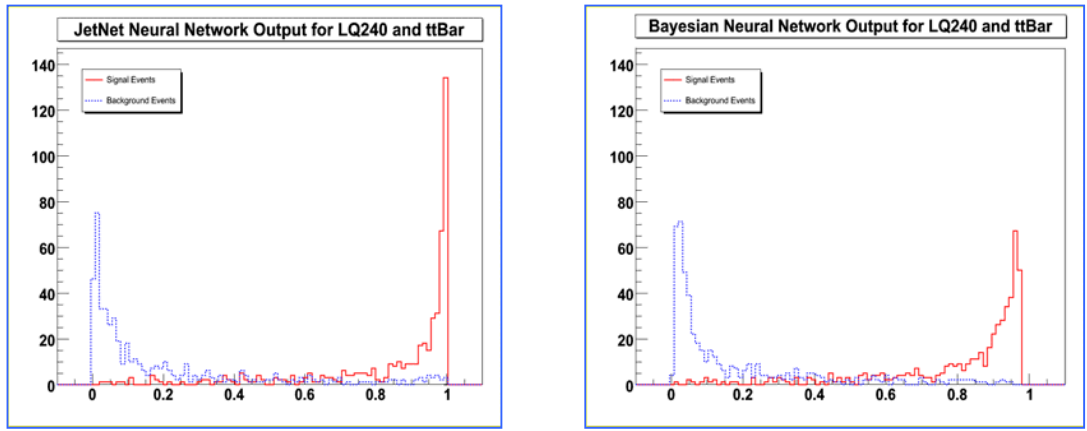

In Fig. 3, the average Bayesian neural network output is compared with the 100 networks that were averaged over. The average can be seen to give better generalization. The discrimination power for the two approaches is compared in Fig. 4. The Bayesian NN approach is seen to yield better signal efficiency for similar background, sometimes up to 35%.

A similar neural network analysis to discriminate against the Drell-Yan process and LQ (mass= 240 GeV) was also performed. The same variables were used in the training and testing of each neural network. The Bayesian neural network provided a significant improvement in signal/background discrimination over the Conventional neural network. However after the cuts were applied only 150 events remained for both signal and background. Those familiar with conventional neural networks will realize that this is not enough data to perform conventional neural network analysis. We are currently performing the analysis by generating a larger sample of Drell-Yan events to verify our findings.

**Figure 3.** Average Bayesian NN output distributions (thick histograms) for signal (red) and $t\bar{t}$ background (blue) superposed on the 100 individual networks.

**Figure 4.** Comparison of signal vs. background for Bayesian (blue dots) and conventional (red dots) NN.

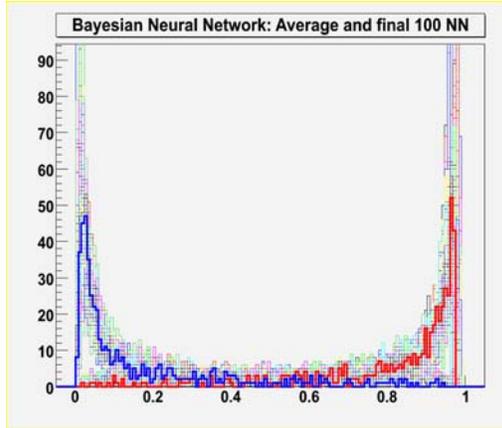
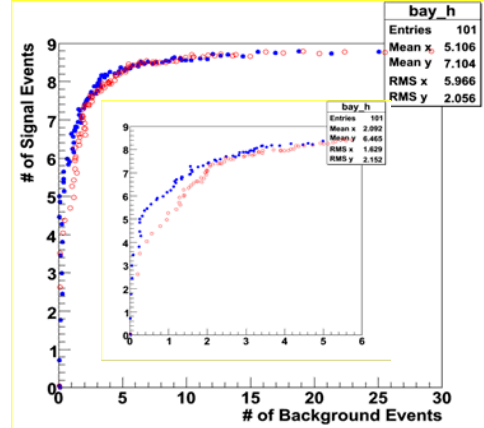

## CONCLUSIONS

Bayesian learning of neural networks provides more optimal and robust results in signal/background discrimination. Better suppression of background in the signal region is found using Bayesian neural networks than with conventional neural networks. The signal efficiency is significantly higher for Bayesian neural networks than for conventional neural networks with small backgrounds. Also, over-fitting is avoided with the Bayesian NN, resulting in much better generalization in training.

## ACKNOWLEDGMENTS

This work is supported in part by a grant from the National Science Foundation. Fermi National Accelerator Laboratory is operated by the Universities Research Association, under a contract, for the U.S. Department of Energy.